\documentclass[preprint,aps,prc,showpacs,nofootinbib]{revtex4}

\usepackage{epsfig}
\usepackage{graphicx}

\begin{document}
\title{High energy inelastic electron hadron scattering, in peripheral kinematics. Sum rules for hadron form factors}

\author{E. A. Kuraev, M. Se\v cansk\'y*}
\affiliation{\it JINR-BLTP, 141980 Dubna, Moscow region, Russian
Federation}

\author{E. Tomasi-Gustafsson}

\affiliation{\it DAPNIA/SPhN, CEA/Saclay, 91191 Gif-sur-Yvette Cedex,
France }

\date{\today}

\begin{abstract}
Relations between differential cross section for inelastic
scattering of electrons on hadrons and hadron form factors (sum
rules) are derived on the basis of analytical properties of heavy
photon forward Compton scattering on hadrons. Sum rules relating the
slope of form-factors at zero momentum transfer and anomalous
magnetic moments of hadrons with some integrals on photo-production
on a hadrons is obtained as well. To provide the convergence of
these integrals we construct differences of individual sum rules for
different hadrons. Universal interaction of Pomeron with nucleons is
assumed. We derive the explicit formulae for processes of
electro-production on proton and light isobar nuclei. Sudakov's
parametrization of momenta, for peripheral kinematics relevant here,
is used. The light-cone form of differential cross sections is also
discussed. The accuracy of sum rules estimated in frames of
point-like hadrons and it is shown to be at the level of precision
achievable by experiments. Suggestions and predictions for future
experiments are also given.

\vspace{3cm}

*){\it On leave of absence from Istitute of Physics SAS,
Bratislava}
\end{abstract}

\maketitle

\section{Introduction}

 The idea of construction of sum rules, relating the form-factors of
electron with the cross sections of electro-production processes
at $e^+e_-$ high energy collisions was born in 1974. In a series
of papers the cross sections of processes such as
$$e_+e_-\to
(2e_-e_+)e_+;(e_-\gamma)e_+;(e_-2\gamma)e_+;(e_-\mu_+\mu_-)e_+$$
were calculated in the so called peripheral kinematics, when the jet
consisted from particles noted in parentheses are moving closely to
the initial electron direction in center of mass reference frame.
These cross sections do not decrease as a function of the center of
mass total beam energy $\sqrt{s}$. Moreover, they are enhanced by a
logarithmical factor $\ln(s/m_e^2)$,which is characteristic for
Weizsacker-Williams approximation. It was obtained for contribution
to the process of muon pair production cross section from the so
called bremsstrahlung mechanism (corresponding to the virtual photon
conversion into muon - anti-muon pair) \linebreak
\cite{Ku74a,Ku74b}:
\begin{equation}
\sigma^{e^+e^-\to e^-\mu^+\mu^-e^+}=\displaystyle\frac{\alpha^4}{\pi\mu^2}
\left [\left( ln \displaystyle\frac{s^2}{m_e^2\mu^2}\right)\left(
\displaystyle\frac{77}{18}\xi_2- \displaystyle\frac{1099}{162} \right )+const\right ]
\label{eq:eq1}
\end{equation}
where $\xi_2=\displaystyle\frac{\pi^2}{6}$.

In the paper of Barbieri,Mignaco and Remiddi \cite{Ba72} from 1972
the slope of the electron Dirac Form Factor for $q^2\to 0$ was
calculated :
\begin{equation}
F_1'(0)=\displaystyle\frac{\alpha^2}{8\pi^2\mu^2}\left(\displaystyle\frac{77}{18}\xi_2- \displaystyle\frac{1099}{162} \right).
\label{eq:eq2}
\end{equation}
The fact that the same coefficients appear in Eqs (\ref{eq:eq1})
and (\ref{eq:eq2}) suggests that a relation exists between
inelastic cross section and elastic form factors. This relation
was later on derived in the paper \cite{Ku74b}.

Here we review an extension of these studies to strong interaction
particles, considering QED interactions in the lowest order of
perturbation theory. We demonstrate sum rules which relate the
nucleon and light nuclei form-factors with the differential cross
sections of electron scattering on the corresponding hadrons in
peripheral kinematics. This kinematics corresponds to the region
of very small values of Bjorken parameter $x_B$ in deep inelastic
scattering experiments. This paper is organized as follows.

After describing the relevant processes in peripheral kinematics,
in terms of Sudakov variables (Section II), we remind the
analytical properties of the advanced and the retarded parts of
the virtual Compton scattering amplitudes (Section II). Then
briefly the problem of restoring gauge invariance and formulation
of the modified optical theorem are discussed.

Following QED analysis in \cite{Ku74b}, we introduce light cone
projection of Compton scattering amplitude integrated on some
contour in the $s_2$ plane, where $s_2$ is the invariant mass
squared of the hadronic jet.

Sum rules (Section III) arise when the Feynman contour in the
$s_2$ plane is closed to the left real axes singularities of the
Compton amplitude and to the right ones. Sum rules obtained in
such a way contain the left hand cut contribution which is
difficult to be interpreted in terms of cross sections. Moreover,
it suffers from ultraviolet divergencies of contour integral
arising from Pomeron Regge pole contribution. The final form of
these sum rules consists of differences, constructed in such a way
in order to compensate the Pomeron contributions and the left hand
cuts as well.

The applications to different kinds of targets, as proton and
neutron, deuteron and light nuclei are explicitly given. Appendix
A is devoted to an estimation of the cross section of $p\bar{p}$
pair proton photo-production in the framework of a simple model
arising from effect of identity of protons in final state.
 In Appendix B  the details of kinematics of recoil
target particle momentum is investigated in terms of Sudakov's
approach.
\subsection{Sudakov parametrization}

\begin{figure}
\begin{center}
\includegraphics[width=6cm]{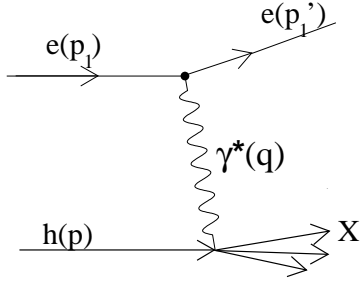}
\caption{\label{Fig:fig1} Feynman diagram for inelastic electron hadron scattering .}
\end{center}
\end{figure}

Let us consider the process presented in Fig. \ref{Fig:fig1},
where the inelastic electron - hadron interaction occurs through a
virtual photon of momentum $q$. The particle momenta are indicated
in the figure: $p^2=M^2$, $p_1^2=p_1'^2=m^2$. The total energy is
$s=(p+p_1)^2$ and the momentum transfer from the initial to the
final electron is $t=(p_1-p_1')^2$.

Let us introduce some useful notations, in order to calculate the
differential cross section for the process of Fig. \ref{Fig:fig1},
where the the hadron is a proton in peripheral kinematics, i.e,
$s\gg -t\simeq M^2$.  Therefore $s=(p_1+p)^2=M^2+m_e^2+2p
p_1\simeq 2pp_1=2ME\gg M^2\gg m_e^2$.

The differential cross section can be written as:
\begin{equation}
d\sigma =\displaystyle\frac{1}{2\cdot 2\cdot 2 s}\sum |{\cal M}|^2d\Gamma.
\label{eq:sig}
\end{equation}

Let us define two light-like vectors:
$$\tilde p =p-p_1 \displaystyle\frac{M^2}{s},~
\tilde p_1 =p_1-p \displaystyle\frac{m^2}{s}$$
and a transversal vector, $q_\perp$, such that $\tilde p q_\perp= p_1q_\perp=0.$

Therefore ${\tilde p}^2={\cal O}\left(\displaystyle\frac
{m^2M^4}{s^2}\right )$ and similarly ${\tilde p}^2={\cal
O}\left(\displaystyle\frac{m^4M^2}{s^2}\right )$. Terms of order
${\cal O}\left(\displaystyle\frac {M^2}{s},\frac {m^2}{M^2} \right
)$ compared to ones of order 1 we will neglect systematically. In
the Laboratory system, with an appropriate choice of the axis, the
four vectors are written, in explicit form, as: $\tilde p_1 \approx
p_1=E(1,1,0,0)$, and $\tilde p=\displaystyle\frac{M}{2}(1,-1,0,0)$,
and $q_{\perp}=(0,0,q_x,q_y)$, $q^2_{\perp}=-\vec q^2<0$, which is
essentially a two-dimensional vector. Let us express the four
momentum of the exchanged photon in the Sudakov parametrization, in
infinite momentum frame, as function of two (small)  parameters
$\alpha$ and $\beta$ (Sudakov parameters):
\begin{equation}
q=\alpha\tilde p+\beta \tilde p_1+q_\perp.
\label{eq:sud}
\end{equation}

The on-mass shell condition for the scattered electron can be written as:
\begin{equation}
p_1'^2-m^2=(p_1-q)^2-m^2=-\vec q^2+s\alpha\beta-\alpha s-\beta m^2=0
\label{eq:eqms}
\end{equation}
where we use the relation:
\begin{equation}
2p_1\tilde p_1=2p_1\left(p_1-p \displaystyle\frac{m^2}{s}\right )=m^2,
\label{eq:eqms1}
\end{equation}
similarly one can find $2p\tilde p=M^2$.

From the definition, Eq. (\ref{eq:sud}), and using Eq.
(\ref{eq:eqms}), the momentum squared of the virtual photon is:
$$q^2=s \alpha\beta -\vec q^2=-\displaystyle\frac{\vec q^2+m^2\beta^2}{1-\beta}<0.$$
The variable $\beta$ is related to the invariant mass of the
proton jet, the set of particles moving close to the direction of
the initial proton:
$$s_2=(q+p)^2-M^2+{\vec q}^2\simeq s\beta$$ neglecting small terms
as $s\alpha\beta$ and $\alpha M^2$ (Weizsacker-Williams
approximation \cite{We34}). We discuss later the consequences of
such approximation on the $s$ dependence of the cross section. In
these notation the phase space of the final particle
\begin{eqnarray}
d\Gamma &=&(2\pi)^4\displaystyle\frac{d^3p_1'}{2\epsilon_1'
(2\pi)^3}\Pi_1^n \displaystyle\frac{d^3q_i}{2\epsilon_i (2\pi)^3}
\delta^4(p_1+p-p_1'-\sum_i^n q_i), \label{eq:eq7}
\end{eqnarray}
introducing an auxiliary integration on the photon transferred
momentum $\int d^4q \delta^4(p_1-q-p_1')=1$ can be written as
\begin{eqnarray}
d\Gamma = (2\pi)^{-3} \delta^4((p_1-q)^2-m^2)d^4q d\Gamma_H,
\end{eqnarray}
where $d\Gamma _H$ is the hadron phase space:
\begin{equation}
d\Gamma _H=(2\pi)^4\delta^4(p +q-\sum_i^nq_i)\Pi_1^n \displaystyle\frac{d^3q_i}{2\epsilon_i (2\pi)^3}.
\label{eq:eq8}
\end{equation}

In the Sudakov parametrization:
\begin{equation}
d^4q=\displaystyle\frac{s}{2} d\alpha d\beta d^2q_\perp\simeq
\displaystyle\frac{ds_2}{2s} d(s\alpha) d^2\vec q, \label{eq:eq9}
\end{equation}
we obtain:
\begin{equation}
d\Gamma=\displaystyle\frac{ds_2}{2s} d^2\vec q (2\pi)^{-3}d\Gamma_H.
\label{eq:eq10}
\end{equation}
Let us use the matrix elements to be expressed by the Sudakov
parameters. Then it can be rewritten into the form:
\begin{equation}
{\cal M}=\displaystyle\frac{4\pi\alpha}{q^2}\overline
u(p_1')\gamma^{\mu}u(p_1) {\cal J}_H^{\nu}g_{\mu\nu}.
\label{eq:eq11}
\end{equation}

It is convenient to use here the Gribov representation of the
numerator of the (exact) Green function in the Feynman gauge for
the exchanged photon:
\begin{equation}
g_{\mu\nu}=(g_{\perp})_{\mu\nu}+\displaystyle\frac{2}{s}(\tilde p_{\mu}\tilde p_{1\nu} +
\tilde p_{\nu}\tilde p_{1\mu}).
\label{eq:eq12}
\end{equation}
All three terms, in the right hand side of the previous equation,
give contributions to the matrix element proportional to
$$1:\displaystyle\frac{s}{M^2}:\displaystyle\frac{M}{s}.$$
So, the main contribution (with power accuracy) is
\begin{equation}
{\cal M}=\displaystyle\frac{4\pi \alpha }{q^2}
\displaystyle\frac{2}{s} {\overline u} (p_1')\hat p u(p_1){\cal
J}_H^{\nu}p_{1\nu}= \displaystyle\frac{8\pi \alpha s}{q^2} N
\displaystyle\frac {{\cal J}_H^{\nu}p_{1\nu}}{s}, \label{eq:eq13}
\end{equation}

with $N=\displaystyle\frac{1}{s} {\overline u}(p_1')\hat p
u(p_1)$. One can see explicitly the proportionality of the matrix
element of peripheral processes to $s$, in the high energy limit,
$s\gg -t$. It follows from the relation
$\sum_{pol}|N|^2=\displaystyle\frac{1}{s^2} Tr \hat p_1'\hat p\hat
p_1\hat p=2$ and from the fact, that the quantity $
\displaystyle\frac{1}{s}{\cal J}_H^{\nu}p_{1\nu}$ is finite in
this limit. Such term can be further transformed using the
conservation of the hadron current ${\cal
J}_H^{\nu}q_{\nu}\simeq(\beta p_1+q_{\perp})_{\nu}{\cal
J}_H^{\nu}=0$. The latter leads to
\begin{equation}
\displaystyle\frac{1}{s}{\cal J}_H^{\nu}
p_{1\nu}=\displaystyle\frac{1}{s\beta} \vec q\cdot \vec {\cal J}_H=
\displaystyle\frac{|\vec q|}{s_2} (\vec e\cdot \vec {\cal J}_H),
\label{eq:eq14}
\end{equation}
where $\vec e=\vec q/|\vec q|$ is the polarization vector of the
virtual photon. As a result, one finds:
\begin{equation}
\sum |{\cal M}|^2= \displaystyle\frac{(8\pi \alpha s)^2|\vec q|^2}
{\left [(\vec q)^2+m^2\left (\displaystyle\frac{s_2}{s}\right )^2
\right ]^2} \displaystyle\frac{2}{s_2^2} \left (\vec {\cal J}_H\cdot
\vec e\right )^2. \label{eq:eq15}
\end{equation}

With the help of Eqs. (\ref{eq:eq7},\ref{eq:eq13},\ref{eq:eq15})
one finds:
\begin{equation}
d\sigma^{(e+p\to e+jet)}= \displaystyle\frac{\alpha^2d^2q\vec q^2
ds_2} {\pi \left [(\vec q)^2+m^2\left
(\displaystyle\frac{s_2}{s}\right )^2\right ]^2s_2^2} (\vec e\cdot
\vec{\cal J}_H)^2 d \Gamma_H. \label{eq:eq17}
\end{equation}

Let us note that the differential cross section at $\vec q^2\ne 0$
does not depend on the CMS energy $\sqrt{s}$. In the logarithmic
(Weizsacker-Williams) approximation, the integral over the
transverse momentum $\vec q$, at small $\vec q^2$, gives even rise
to large logarithm:
\begin{equation}
\sigma_{tot}^{\ell}= \pi \alpha ~ln \left( \displaystyle\frac{s^2
Q^2} {M^4m^2}\right ) \int_{s_{th}}^\infty
\displaystyle\frac{ds_2}{s_2}\sigma_{tot}^{(\gamma+p\to X)}(s_2),
\label{eq:qq3}
\end{equation}
where $s_{th}=(M+m_{\pi})^2-M^2$, $Q^2$ is the characteristic momentum transfer
squared, $Q^2\simeq M^2$, and we introduced the total cross section for real
polarized photons interacting with protons:
\begin{equation}
\sigma_{tot}^{(\gamma+p\to X)}(s_2,q^2=0)=\displaystyle\frac{\alpha\pi}{s_2}\int
(\vec {\cal J}_H\cdot \vec e)^2 d \Gamma_H.
\label{eq:qq4}
\end{equation}

\begin{figure}
\begin{center}
\includegraphics[width=14cm]{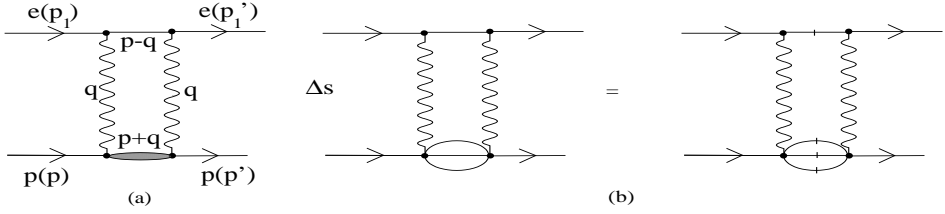}
\caption{\label{Fig:fig2} Feynman diagram for $ e^+e^-$ scattering at the order $\alpha^3$.}
\end{center}
\end{figure}
The differential cross section $(\ref{eq:qq3})$ is closely related (due to the optical
theorem) with the $s$-channel discontinuity of the forward amplitude for electron-proton
scattering with the same intermediate state: a single electron and a jet, moving in
opposite directions (see Figs. \ref{Fig:fig2}a, \ref{Fig:fig2}b) where, by Cutkovsky
rule, the denominators of the "cutted" lines in the Feynman graph of Fig. \ref{Fig:fig2}b
must be replaced by:
\begin{equation}
\displaystyle\frac{1}{q^2-M^2+i0}\to -2\pi i\delta(q^2-m^2).
\label{eq:qq4a}
\end{equation}
For the spin averaged forward scattering amplitude we have:
\begin{equation}
\Delta_sA(s)=\displaystyle\frac{4s\alpha}{\pi^2}
\int \displaystyle\frac{d^2q_{\perp}\vec q^2}{(q^2)^2}
\int \displaystyle\frac{ds_2}{s_2}\sigma^{(\gamma^*p\to X)}(s_2,q)
\label{eq:qq5}
\end{equation}
with
\begin{equation}
\sigma^{(\gamma^*p\to X)}(s_2,q)=\int \displaystyle\frac{4 \pi
\alpha}{2\cdot 2\cdot 2s_2}(\vec {\cal J}_H\cdot \vec e)^2 d
\Gamma_H. \label{eq:qq5a}
\end{equation}
From the formulae given above we can obtain
\begin{equation}
\frac{d\Delta_s\bar{A}^{eY\to eY}}{d^2q}=2s\frac{d\sigma^{eY\to
eY}}{d^2q},
\end{equation}
where $\bar{A}^{eY\to eY}$ is the averaged on spin states forward
scattering amplitude. This relation is the statement of optical
theorem in differential form.

Let us now consider the discontinuity of forward scattering
amplitude with the initial hadron intermediate state, we call it a
"pole contribution". For the case of elastic electron-proton
scattering we have
\begin{equation}
\frac{d\Delta_s\bar{A}^{ep\to
ep}}{d^2q}=\frac{(4\pi\alpha)^2}{(q^2)^2s(2\pi)^2}Sp,
\end{equation}
with
$$Sp=Sp(\hat{P}+M_p)\Gamma(q)(\hat{P}'+M_p)\Gamma(-q)^*,$$
and
$$\Gamma(q)=F_1\hat{p}_1-\frac{1}{2M}F_2\hat{q}\hat{p}_1.$$

A simple calculation gives $Sp=2s^2[(F_1)^2+\tau(F_2)^2]$, with
$\tau=\vec{q}^2/(4M^2_p)$.

For the case of electron-deuteron scattering we use the
electromagnetic vertex of deuteron in the form \cite{AR78}
\begin{equation}
<\xi^{\lambda'}(P')|J^{EM}_\mu(q)|\xi^\lambda(P)>=d_\mu[F_1(\xi^{\lambda'*}\xi^\lambda)-
\frac{F_3}{2M_d^2}(\xi^{\lambda'*}q)(\xi^\lambda
q)]+F_2[\xi^\lambda_\mu(q\xi^{\lambda'*})-\xi^{\lambda'*}_\mu(\xi^\lambda
q)],
\end{equation}
where $d_\mu=(P'+P)_\mu,q_\mu=(P'-P)_\mu $ and $\xi^\lambda(P)$ is
the polarization vector of deuteron in chiral state $\lambda$. It
has the properties:
\begin{equation}
\xi^2=-1,(\xi(P)P)=0;\sum_\lambda
\xi^\lambda(P)_\mu\xi^{\lambda*}(P)_\nu=g_{\mu\nu}-\frac{P_\mu
P_\nu}{M_d^2}.
\end{equation}
For the averaged on spin states forward scattering amplitude we
have
\begin{equation}
\frac{d\Delta_s\bar{A}^{ed\to
ed}}{d^2q}=\frac{2s(4\pi\alpha)^2}{3(q^2)^2(2\pi)^2}Tr,
\end{equation}
with
\begin{equation}
Tr=2(F_1)^2+(F_1+2\tau_d(1+\tau_d)F_3)^2+2\tau_d(F_2)^2,\tau_d=\frac{\vec{q}^2}{4M_d^2}.
\end{equation}
We note that the amplitude corresponding to box-type Feynman
diagram with the crossed photon legs have a zero $s$-channel
discontinuity.
\section{Virtual Compton scattering on proton}

Let us examine the different contributions to the total amplitude
for virtual photon Compton scattering on a proton (hadron).
Keeping in mind the  baryon number conservation low we can
separate all possible Feynman diagram to four classes. In one,
which will be named as class of retarded diagram (corresponding
amplitude is denoted as $A_1$) the initial state photon is first
absorbed by nucleon line and after emitted the scattered photon.
Another class (advanced, $A_2$), corresponds to the diagrams, in
which scattered photon is first emitted along the nucleon line and
the point of absorption is located after. Third class corresponds
to the case when both photons do not interact with initial nucleon
line. The corresponding amplitude is which we denote as $A_P$. The
fourth class contains diagrams in which only one of external
photons interact with nucleon line. The corresponding notation is
$A_{odd}$ (see Fig.3):
\begin{equation}
A^{\mu\nu}(s,q)=A_1^{\mu\nu}(s,q)+ A_2^{\mu\nu}(s,q)+
A_{P}^{\mu\nu}(s,q)+A_{odd}. \label{eq:amp}
\end{equation}
Amplitude $A_{P}^{\mu\nu}(s,q)$ corresponds to Pomeron type
Feynman diagram (Fig. \ref{Fig:fig4}e) and gives the non-vanishing
contribution to the total cross sections in the limit of a large
invariant mass squared of initial particles $s_2\to \infty$. The
fourth class amplitude can be relevant in experiments with
measuring the charge-odd effects. It is not considered here. One
can show explicitly that each of 4 classes amplitudes are gauge
invariant. The arguments in favor of it is essentially the same as
was used in QED case \cite{BFKK}.
\begin{figure}
\begin{center}
\includegraphics[width=16cm]{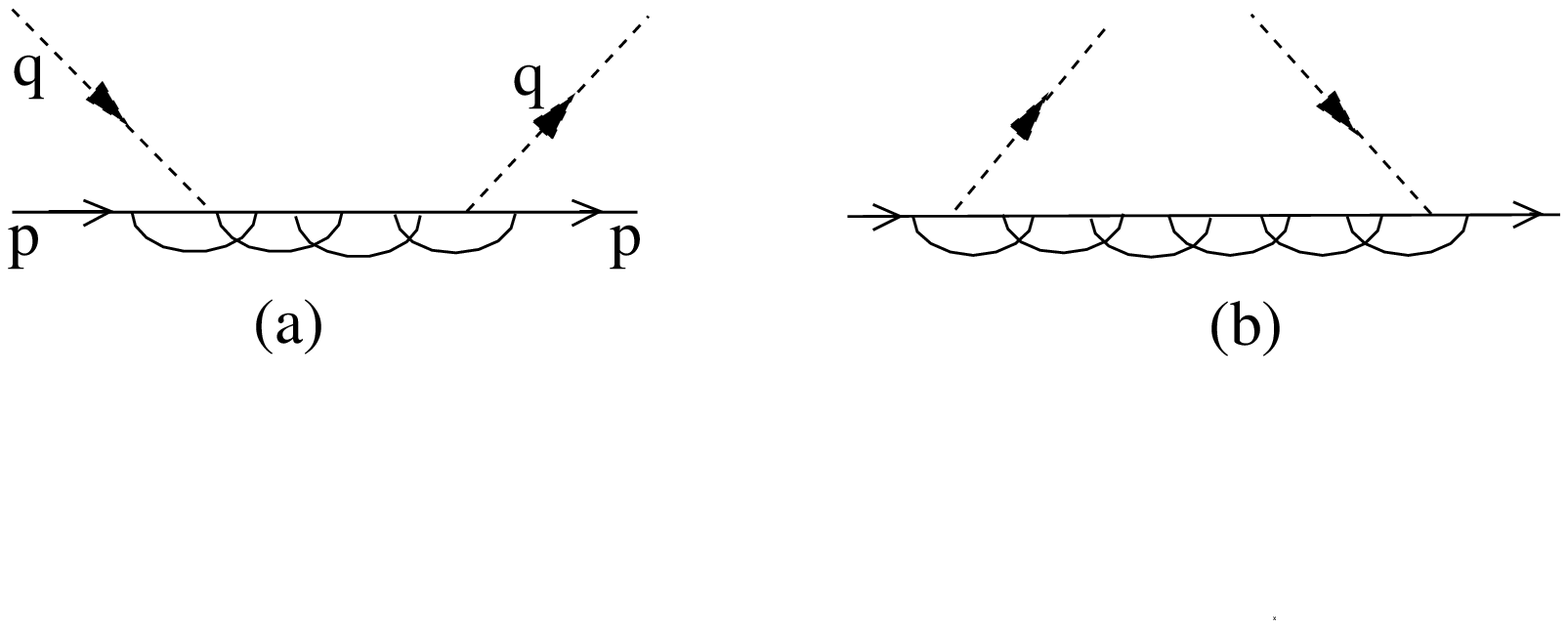}
\caption{\label{Fig:fig3} Illustration of retarded  (a) and advanced (b)
virtual photon emission and absorbtion diagrams. The diagram containing
Pomeron is not considered here.}
\end{center}
\end{figure}

 Let us discuss now the analytical properties  of the
retarded part of the forward Compton scattering of a virtual
photon on a proton, $A_1(s_2,q)$ (see Fig. \ref{Fig:fig4})at $s_2$
- plane. Due to general principles the singularities - poles and
branch points - are situated on the real axis.

These singularities are illustrated in Fig. \ref{Fig:fig5}. On the
right side the pole at $s_2=0$ correspond to one nucleon exchange
in the $s_2$ channel (Fig. \ref{Fig:fig5}a), the right hand cut
starts at the pion-nucleon threshold, $s_2=(M+m_{\pi})^2-M^2$. The
left cut, related with the $u$ channel 3-nucleon state  of the
Feynman amplitude is illustrated in Fig. \ref{Fig:fig4}f. It is
situated rather far from the origin at $s_2=-8M^2$. It can be
shown that it is the nearest singularity of right hand cut. Really
the $u$-channel cut corresponding to $2\pi N$ state can not be
realized without exotic quantum number states (see Fig.
\ref{Fig:fig4}h).

\section{Sum rules}

Following to \cite{BFKK} let us introduce the quantity
\begin{equation}
\int_C ds_2\displaystyle\frac{p_1^{\mu}p_1^{\nu}
A_{1\mu\nu}^{\gamma^*p\to\gamma^*p}}{s^2({\vec q}^2)^2}=
\displaystyle\frac{dI}{d{\vec q}^2} \label{eq:amp1},
\end{equation}
with the Feynman contour $C$ in the $s_2$ plane as it is shown in
Fig. 5a.

 Sum rules appear when one considers the equality of the
path integrals along the contours obtained by deformation $C$ such
to be closed to the left and to the right side (Fig.
\ref{Fig:fig5}). As a result one finds:
\begin{equation}
F({\vec q}^2)=\displaystyle\frac{d\sigma_{left}}{d{\vec q}^2}-\displaystyle\frac{d\sigma_{el}
-d\sigma_{el}^B}{d{\vec q}^2}
=\displaystyle\frac{d\sigma_{inel}}{d{\vec q}^2},
\label{eq:sr1}
\end{equation}
where $\displaystyle\frac{d\sigma_{left}}{d{\vec q}^2}$ indicates
the contribution of left cut \footnote{The coincidence of numbers in
(1,2) derives from the absence of left cut contribution, which is
known for planar Feynman diagram amplitudes.};
$$\displaystyle\frac{d\sigma_{el}^B}{d{\vec q}^2}=\displaystyle\frac{4\pi\alpha^2 Z^2}
{({\vec q}^2)^2}.$$ The latter is generally the Born cross section
of the scattering of an electron on any hadron with charge $Z$, when
the strong interaction is switched off, and
$\displaystyle\frac{d\sigma_{el}}{d{\vec q}^2}$ is the elastic
electron hadron cross section, when the strong interaction is
switched on, in the lowest order of QED coupling constant. This
quantity can be expressed in terms of electromagnetic form factors
of corresponding hadrons.

Using the notation for generalized square of form-factors as
$\Phi^2$ we have for process of electron scattering on a hadron
$Y$ with charge $Z$ the expression:
\begin{equation}
Z^2-\Phi^2(-\vec{q}^2)=\frac{2\vec{q}^2}{\pi\alpha^2}\frac{d\sigma^{eY\to
eX}}{d\vec{q}^2}.
\end{equation}
For the case of spin-zero target the quantity $\Phi^2$ coincide
with its squared charge form factor.

For the case of electron scattering on a spin one-half (proton,
$^3\!He$,$^3\!H$),which are described by two form-factors (Dirac's
one $F_1$ and Pauli one $F_2$) we have
\begin{equation}
Z_i^2-F_{1i}^2(-\vec{q}^2)-\tau_i
F_{2i}^2(-\vec{q}^2)=\frac{2\vec{q}^2}{\pi\alpha^2}\frac{d\sigma^{eY_i\to
eX_i}}{d\vec{q}^2},
\end{equation}
with
$$\tau_i=\frac{\vec{q}^2}{4M_i^2};Z_p=Z_{^3\!H}=1;Z_{^3\!He}=2. $$

For scattering of electron on deuteron we have:
\begin{equation}
1-\frac{1}{3}[2F_1^2(-\vec{q}^2)+
[F_1(-\vec{q}^2)+2\tau_d(1+\tau_d)F_3(-\vec{q}^2)]^2+ 2\tau_d
F_2^2(-\vec{q}^2)]=\frac{2\vec{q}^2}{\pi
\alpha^2}\frac{d\sigma^{ed\to eX}}{d\vec{q}^2}.
\end{equation}
 These equations can be tested in experiments with electron-hadron
 colliders.

Further we will consider differential form of these sum rules: $
\left .\displaystyle\frac{d}{d{\vec q}^2}({\vec q}^2)^2F({\vec
q}^2) \right |_{{\vec q}^2=0}$, which can be expressed in terms of
charge radii of hadrons, their anomalous magnetic moments, etc.
and photo-production total cross section.

Considering formally this derivative at $\vec{q}^2=0$ we obtain
\begin{equation}
\frac{d}{d\vec{q}^2}\Phi_Y^2|_{\vec{q}^2=0}=
\frac{2}{\pi^2\alpha}\int\limits_{s_{th}}^\infty\frac{ds_2}{s_2}\sigma_{tot}^{\gamma
Y\to X}(s_2).
\end{equation}
Unfortunately the sum rule in this form can not be used for
experimental verification due to the divergence of integral in
right hand side of this equation. Its origin follows from the
known fact of increasing of photo production cross sections at
large values of initial center of mass energies squares $s_2$.
It's commonly known that this fact is the consequence of Pomeron
Regge pole contribution. The universal character of Pomerom
interaction with nucleons can be confirmed by the Particle Data
Group-2004 (PDG) result
\begin{equation}
[\sigma^{\gamma p}(s_2)-\sigma^{\gamma n}(s_2)]|_{s_2\to
\infty}=[2\sigma^{\gamma p}(s_2)- \sigma^{\gamma d}(s_2)]|_{s_2\to
\infty}=0.
\end{equation}

In paper \cite{Ba04} the difference of proton and neutron sum rule
was derived
\begin{equation}
\displaystyle\frac{1}{3} <r^2_p>+ \displaystyle\frac{1}{4M^2}\left
[\kappa^2_n-\kappa^2_p\right ]= \displaystyle\frac{2}{\pi^2\alpha}
\int_{\omega_n}^{\infty }\displaystyle\frac{d\omega}{\omega} \left
[\sigma^{\gamma p\to X}(\omega)-\sigma^{\gamma n\to X}(\omega)\right
]. \label{eq:eq23}
\end{equation}
We use here the known relations
$$F_1(-\vec{q}^2)=1-\frac{1}{6}\vec{q}^2<r^2>+O((\vec{q}^2)^2);F_2(0)=\kappa,$$
with $<r^2>, \ \kappa$-are the charge radius squared and the
anomalous magnetic moment of nucleon (in units $h/Mc$).

It was verified in this paper that this sum rule is fulfilled within
the experimental errors: both sides of the equation equal 1.925 mb.
Here the Pomeron contribution is compensated in the difference of
proton and neutron total cross photo-production cross sections.

In paper \cite{Du06} the similar combination of cross sections was
considered  for A=3 nuclei:
\begin{equation}
\displaystyle\frac{2}{3} <r^2_{^3\!He}>-\displaystyle\frac{1}{3}
<r^2_{^3\!H}>- \displaystyle\frac{1}{4M^2}\left
[\kappa^2_{^3\!He}-\kappa^2_{^3\!H}\right ]=
\displaystyle\frac{2}{\pi^2\alpha} \int_{\omega_{th}}^{\infty
}\displaystyle\frac{d\omega}{\omega} \left [\sigma^{\gamma ^3\!He\to
X}(\omega)-\sigma^{\gamma ^3\!H\to X}(\omega)\right ].
\label{eq:eq23a}
\end{equation}

In the similar way, the combination of cross sections of electron
scattering on proton and deuteron leads to the relation
\begin{eqnarray}
\frac{1}{3}<r^2_d>-\frac{F_3(0)}{3M_d^2}-\frac{1}{6M_d^2}F_2(0)^2-
2[\frac{1}{3}<r^2_p>-\frac{1}{4M_p^2}\kappa_p^2]= \nonumber \\
\frac{2}{\pi^2\alpha}\int\limits_{\omega_{th}}^\infty\frac{d\omega}{\omega}
[\sigma_{tot}^{\gamma d\to X}(\omega)-2\sigma_{tot}^{\gamma p\to
X}(\omega)],
\label{eq:eq39}
\end{eqnarray}
with $\omega_{th}$ for deuteron and proton are different:
$(\omega_{th})_d=2,2MeV,(\omega_{th})_p=
m_{\pi}+\frac{m_{\pi}^2}{2M_p} \approx 140MeV. $

  We use here the similar expansion for $F_1$
deuteron form factor $F_1(0)=1$ and introduce its square charge
radius. Other quantities can be found in paper \cite{GM}:
$$F_2(0)=-\frac{M_d}{M_p}\mu_d; \ \mu_d=0.857; \
2F_3(0)=1+F_2(0)-M_d^2Q_d; \
 Q_d=0.2859 fm^2.$$

We find for the left side of formula (\ref{eq:eq39})
by using PDG-2000 and \cite{rossi}
\begin{equation}
\frac{2}{\pi^2\alpha}\biggl\{
\int\limits_{0.020}^{0.260}
\frac{d\omega}{\omega}\sigma_{tot}^{\gamma d\to X}(\omega)
+\int\limits_{0.260}^{16}\frac{ d\omega}{\omega}
\biggl[
\sigma_{tot}^{\gamma d\to X}(\omega)-2\sigma_{tot}^{\gamma p\to X}(\omega)
\biggr]
\biggr\}
=0.8583 fm^2=8.5834 mb
\end{equation}

\section{Conclusions}

The left cut contribution has no direct interpretation in terms of
cross section. In analogy with QED case it can be associated with
contribution to the cross section of process of proton antiproton
pair electro-production on proton $ep\to e{2p\bar{p}}$, arising
from taking into account the identity of final state protons.

Fortunately the threshold of this process is located enough far
away. Using this fact we can estimate its contribution in the
framework of QED- like model with nucleons and pions
($\rho$-mesons),omitting the form factor effects (so we put them
equal to coupling constants of nucleons with pions and vector
mesons).

In this paper we use optical theorem, which connects the s-channel
discontinuity of forward scattering amplitude with the total cross
section. This statement is valid for complete scattering
amplitude, nevertheless, we consider only part of it, $A_1$. We
can explicitly point out on the Feynman diagram (see Fig. 4 g),
contributing to $A_2$, which has 3 nucleon $s$-channel state. The
relevant contribution can be interpreted as identity effect of
proton photo-production of $p \bar{p}$ pair.

The explicit calculation in the framework of our approach is given
by Appendix A.
 The corresponding contribution to the derivative on $\vec{q}^2$ at $\vec{q}^2=0$ of scattering
 amplitudes entering the sum rules have an order of magnitude
 $$ I= \frac{g^4M^2}{\pi^3s_{2 min}^2}.$$
 In order to estimate the strong coupling constant we use the PDG value for
 the total cross section of scattering of pion on proton
 $\sigma_{tot}^{\pi p}=20 mb $. Keeping in mind the $\rho$-meson
 t-channel contribution  $\sigma^{\pi p}\sim g^4/(4\pi m_\rho^2)$
 and the minimal value of three nucleon invariant mass squared
 $s_{2min}=8M_p^2$ we have $I\approx (1/15)mb$. Comparing this
 value with typical values of right and left hand sides of sum
 rules of order of $2 mb$ we estimate the error arising by
 omitting the left cut contribution as well as replacing our
 non-complete cross section by the measurable ones on the level of
 $3\%$.

Using the data and PDG (2000) for $\gamma p$ and $\gamma d$ cross
sections \cite{rossi} we find
$$<r_d> \ \approx \ 1.94 fm.$$
This quantity in a satisfactory agreement with prediction of models based
dispersion relations \cite{d} where is $<r_d> \sim 2fm.$ Reason
(besides the errors 3\% of our approach) can be related with the
lack of data for $\gamma p$ and $\gamma d$ cross sections near the
thresholds.

\begin{figure}
\begin{center}
\includegraphics[width=16cm]{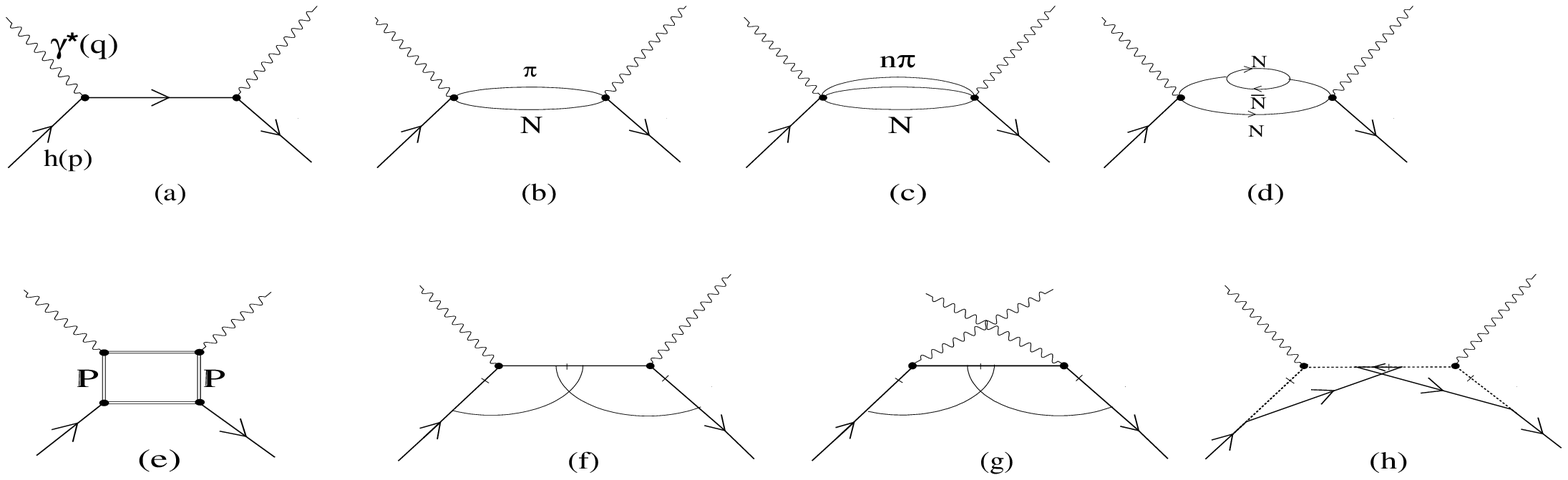}
\caption{\label{Fig:fig4} Feynman diagrams for forward virtual Compton scattering on the proton, contributing to the retarded part of the amplitude: intermediate state in $s_2$ channel for single proton (a), $N\pi$ (b), $Nn\pi$ (c), $\sum N\overline{N}$ (d), two jets $s_2$ channel state, with two Pomeron $t$ channel state (d), $3N$ intermediate state in $u$ channel (e), $3N$ intermediate state in $u_2$ channel (f). The Feynman diagram of the $A_2$ set which has $s$ channel discontinuity is illustrated in (g) and an example of exotic $u$ channel state in (h).}
\end{center}
\end{figure}

\begin{figure}
\begin{center}
\includegraphics[width=12cm]{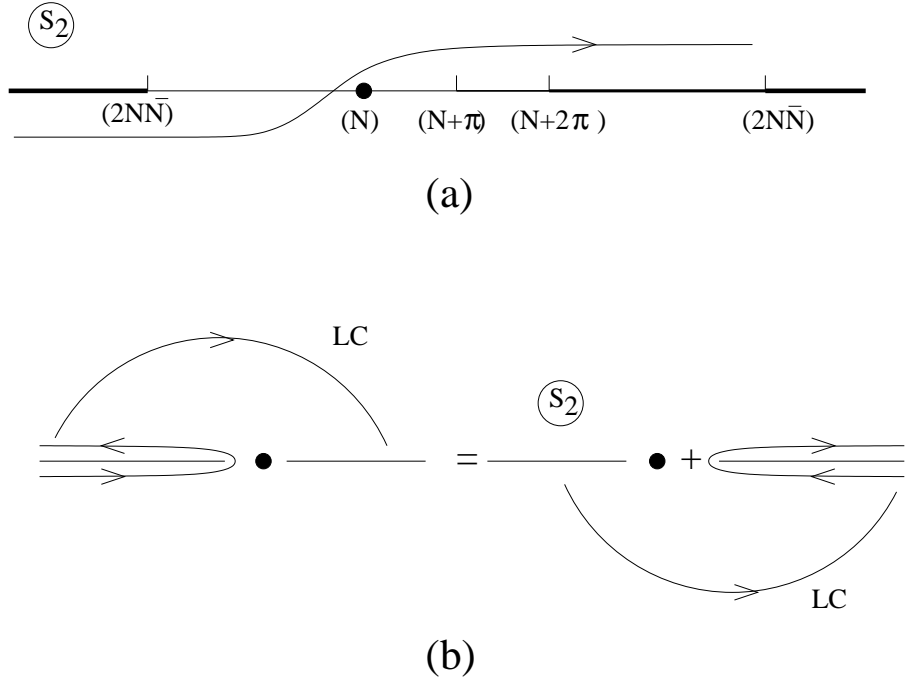}
\caption{\label{Fig:fig5} Illustration of singularities along the $s_2$ real axis with the open contour C (a), and with the contour C closed (b), corresponding o Fig. 4. LC stays for large circle contribution.}
\end{center}
\end{figure}

\section{Acknowledgement}
The authors are thankful to S.Dubnicka for careful reading of the
manuscript and valuable discussions. One of us (E.K.) is grateful
to the Institute of Physics, Slovak Academy of Sciences,
Bratislava for warm hospitality and (M.S.) is grateful to Slovak
Grant Agency for Sciences VEGA for partial support, Gr.No.
2/4099/25. One of us (E.K) is grateful to Saclay Physical Centrum,
Paris, where part of this work was done.
We are grateful to A. V. Shebeko (KHFIT, Kharkov, Ukraina) and V. V. Burov (JINR) for detailed information
about deuteron disintegration region kinematics.

\section{Appendix A:effect of identity of protons to the cross section
of $2p\bar{p}$ photo-production cross section}

The contribution to s-channel discontinuity  of part
$A_2^{\gamma^*p\to\gamma^*p}$ scattering amplitude arising from
interference of amplitudes of creation of proton-antiproton pair
bremsstrahlung type due to identity of protons in the final state
have a form:
\begin{equation}
\Delta_sA_2(s,q)=-\frac{16g^4}{s^2}\int ds_2
d\Gamma_3\frac{S}{(q_1^2-m_\pi^2)(q_2^2-m_\pi^2)},
\end{equation}
where $q_1=P_1-P,q_2=P_2-P$ and
$$
S=\frac{1}{4}Sp(\hat{P}+M)\gamma_5(\hat{P}_1+M)V_1(\hat{P}_3-M)V_2
(\hat{P}_2+M)\gamma_5, $$ and
$V_1=\gamma_5\frac{\hat{q}-\hat{P}_3+M}{d_3}\hat{p}_1+
\hat{p}_1\frac{\hat{P}_1-\hat{q}+M}{d_1}\gamma_5 $; \nonumber \\
$V_2=\gamma_5\frac{-\hat{q}+\hat{P}_2+M}{d_2}\hat{p}_1+
\hat{p}_1\frac{-\hat{P}_3+\hat{q}+M}{d_3}\gamma_5 $; \nonumber \\
$d_{1,2,3}=(q-P_{1,2,3})^2-M^2, s_2=(P+q)^2-M-q^2$.

Besides the phase volume element have a form:
$$
d\Gamma_3=(2\pi)^4\delta^4(P+q-P_1-P_2-P_3)\Pi_1^3\frac{d^3P_i}{2E_i(2\pi)^3}.$$

Here we consider pions to be interacting with nucleons with
coupling constant $g$. The similar expression can be written for
the case when one or both pions are replaced by $\rho$-meson. It
can be shown that the corresponding contributions are
approximately one order of magnitude smaller than those from
pions.

We use Sudakov's parametrization of momenta:
\begin{equation}
q=\alpha\tilde{P}+\beta p_1+q_\bot; \ P=\tilde{P}+\frac{M^2}{s}p_1;
\ P_i=\alpha_i \tilde{P}+\beta_i p_1+P_{i\bot}.
\end{equation}
Using the formulae given above all the relevant quantities can be
written as:
\begin{equation}
\int ds_2d\Gamma_3=\frac{1}{4(2\pi)^3}\frac{d^2P_1
d^2P_2d\alpha_1d\alpha_2}{\alpha_1\alpha_2\alpha_3};
\alpha_1+\alpha_2+\alpha_3=1;\vec{P}_1+\vec{P}_2+\vec{P}_3=\vec{q}
\end{equation}
and
\begin{eqnarray}
s_2=s\beta=-M^2+\sum_1^3\frac{\vec{P}_i^2+M^2}{\alpha_i};
\nonumber \\
q_i^2=-\frac{\vec{P}_i^2+(1-\alpha_i)^2}{\alpha_i}, \ i=1,2; \
d_i=-s_2\alpha_i+2\vec{q}\vec{P}_i-\vec{q}^2, \ i=1,2,3.
\end{eqnarray}
Note that the quantities $V_i$ can be written as
\begin{equation}
V_1=s\gamma_5A_{13}+\frac{\gamma_5\hat{q}\hat{p}_1}{d_3}-\frac{\hat{p}_1\hat{q}\gamma_5}{d_1};
\
V_2=s\gamma_5A_{23}-\frac{\gamma_5\hat{q}\hat{p}_1}{d_2}+\frac{\hat{p}_1\hat{q}\gamma_5}{d_3},
\end{equation}
with
$$A_{13}=\frac{\alpha_1}{d_1}-\frac{\alpha_3}{d_3}; \ A_{23}=
\frac{\alpha_2}{d_2}-\frac{\alpha_3}{d_3}.$$ In this form the gauge
invariance of contribution to forward scattering amplitude is
explicitly seen, namely this quantity turns out to zero at
$\vec{q}\to 0$, (we see that replacements
$\hat{q}\hat{p}_1=\hat{q}_{\perp}\hat{p}_1,
\hat{p}_1\hat{q}=\hat{p}_1\hat{q}_{\perp}$ in $V_{1,2}$ can be
done).

Computation of the trace give the result:
\begin{eqnarray}
\frac{S}{s^2}=A_{13}A_{23}S_1+A_{13}[\frac{1}{d_3}+\frac{1}{d_2}]S_2+
A_{23}[\frac{1}{d_3}+\frac{1}{d_1}]S_3+ [\frac{1}{d_3}+
\frac{1}{d_1}][\frac{1}{d_3}+ \frac{1}{d_2}]S_4,
\end{eqnarray}
with
\begin{eqnarray*}
S_1=M^4+\frac{1}{2}M^2\vec{q}^2+(PP_1)(P_2P_3)+(P_3P_1)(P_2P)-(PP_3)(P_2P_1);
\end{eqnarray*}
\begin{eqnarray*}
S_4=-\frac{\alpha_3\vec{q}^2}{2}[M^2\alpha_3+\alpha_1(PP_2)+\alpha_2(PP_1)-(P_1P_2)];
\end{eqnarray*}
\begin{eqnarray}
S_2=\frac{M^2}{2}[\alpha_2(\vec{q}\vec{P}_2)-\alpha_3(\vec{q}\vec{P}_3)-
(\alpha_1+2\alpha_3)(\vec{q}\vec{P}_1)]+
\frac{1}{2}[(\vec{q}\vec{P}_3)[(P_1P_2)-\\\nonumber
\alpha_2(PP_1)-\alpha_1(PP_2)]+ (\vec{q}\vec{P}_2)[(P_1P_3)+
\alpha_3(PP_1)-\alpha_1(PP_3)]-\\
(\vec{q}\vec{P}_1)[(P_3P_2)-\alpha_2(PP_3)-\alpha_3(PP_2)]];\nonumber
\end{eqnarray}
\begin{eqnarray}
S_3=\frac{M^2}{2}[\alpha_3(\vec{q}\vec{P}_3)-\alpha_1(\vec{q}\vec{P}_1)+\nonumber
(\alpha_2+2\alpha_3)(\vec{q}\vec{P}_2)]+ \frac{1}{2}
[(\vec{q}\vec{P}_1)[-(P_3P_2)-\\\nonumber
\alpha_3(PP_2)+\alpha_2(PP_3)]+ (\vec{q}\vec{P}_2)[(P_1P_3)-
\alpha_3(PP_1)-\alpha_1(PP_3)]-\\\nonumber
(\vec{q}\vec{P}_3)[(P_1P_2)-\alpha_1(PP_2)-\alpha_2(PP_1)]].\nonumber
\end{eqnarray}
The invariants entering this expression have a form
$$2PP_i=\frac{\vec{P}_i^2+M^2(1+\alpha_i^2)}{\alpha_i}; \
2P_iP_j=\frac{(\alpha_i\vec{P}_j-\alpha_j\vec{P}_i)^2+M^2(\alpha_i^2+\alpha_j^2)}{\alpha_i\alpha_j}.
$$

Numerical integration of this expression confirm the estimate
given above within $10\%$.

\section{Appendix B:Correlation between momentum and the
scattering angle of recoil particle in Lab.Frame}

The idea of expanding four-vectors of some relativistic problem
using as a basis two of them (Sudakov's parametrization) becomes
useful in many regions of quantum field theory. It was crucial at
studying the double logarithmical asymptotic of amplitudes of
processes with large transversal momenta. Being applied to
processes with peripheral kinematics it essentially coincides with
infinite momentum frame approach.

Here we demonstrate its application to to study of the kinematic of
peripheral process of jet formation on a resting target particle.
One of experimental approach to study them is measuring the recoil
particle momentum distribution. For instance this method used in
process of electron-positron pair production by linearly polarized
photon on electron in solid target (atomic electrons). Here the
correlation between recoil momentum-initial photon plane and the
plane of photon polarization is used to determine the degree of
photon polarization \cite{VK72}.

Sudakov's parametrization allows to give a transparent explanation
of correlation between the angle of emission of recoil target
particle of mass $M$ with recoil momentum value in laboratory
reference frame \cite{VK72}:
\begin{equation}
\frac{\vec{P}'}{M}=\frac{2\cos\theta_p}{\sin^2\theta_p};\frac{E'}{M}=\frac{1+\cos^2\theta_p}{\sin^2\theta_p},
\end{equation}
where $\vec{P}',E'$are the 3-momentum and energy of recoil
particle,$(E')^2-(P')^2=M^2$; $\theta_p$ is the angle between the
beam axes $\vec{k}$ in the rest frame of target particle
$$\gamma(k)+P(P)\to jet+P(P'), \ s=2kP=2M \omega,P-P'=q, \
P^2=(P')^2=M^2.$$ We consider here the kinematics of main
contribution to the cross section, which correspond the case when
the jet move close to projectile direction. Using Sudakov
representation for transfer momentum $q=\alpha \tilde{P}+\beta
k+q_\bot$, and the recoil particle on mass shell condition
$(P-q)^2-M^2\approx -s\beta-\vec{q}^2=0$, we obtain for the ratio of
squares of transversal and longitudinal components of the 3-momentum
of recoil particle:
\begin{equation}
\tan^2\theta_p=\frac{\vec{q}^2}{(\beta\omega)^2}=\frac{4M^2}{\vec{q}^2},
\vec{q}^2=(P')^2\sin^2\theta_p.
\end{equation}
The relation noted in beginning of this section follows
immediately.

This correlation was first mentioned in paper \cite{BM} where the
production on electrons from the matter was investigated. It was
proven in paper \cite{VK72}.

This relation can be applied in experiments with collisions of high
energy protons scattered on protons in the matter.

\end{document}